\documentclass{article}
\usepackage{a4,amsmath,amsfonts}
\usepackage{graphics}
\usepackage{graphicx}

\usepackage{tikz}
\usepackage{slashed}
\usepackage{fancyhdr}
\usepackage{setspace}

\allowdisplaybreaks

\begin{document}
\begin{center}
\textbf{\huge Kaon Structure in the Confining Nambu-Jona-Lasinio Model}
\vskip 0.3cm \Large Parada.~T.~P.~Hutauruk$^a$,
\\[0.5cm]
\begin{spacing}{1}
\small
\vskip 0.1cm $^a$ \textit{Asia Pacific Center for Theoretical Physics, Pohang, Gyeongbuk 37673, South Korea}\\
\vskip 0.2cm
\end{spacing}
\end{center}

\noindent \textbf{Abstract.} The elastic electromagnetic form factors in the space-like region and valence quark distribution functions for the $K^{+}$ meson are calculated using the confining Nambu-Jona-Lasinio model with the help of the proper-time regularization scheme, which simulates quark confinement. In this model framework, the dynamics information on the nonperturbative aspects are obtained from quark propagators and bound state amplitudes via Bethe-Salpeter equations. We found that the results on the kaon form factors and valence quark distribution functions are qualitatively in excellent agreement with the existing kaon data as well as the perturbation QCD prediction at higher $Q^2$. 
\\
\noindent {\bf Keywords:} Kaon form factors, Kaon parton distribution functions, Nambu-Jona-Lasinio Model.

\begin{center}
\textbf{I. INTRODUCTION}
\end{center}

It is widely known that the kaon consist of a quark-antiquark pair. This shows the structure of the kaon is simpler than the nucleon and hence the dynamics of quarks inside the kaon may be easier to study than that of the nucleon. This gives us a great opportunity to gain useful information of the dynamics of strange quark inside the kaon~[1, 2, 3, 4, 5, 6]. In addition, this may eventually be applied to study the dynamics of quarks inside the nucleon~[5] and to understand the QCD as the underlying theory~[7]. 

In this work the kaon structure is studied by means of the kaon elastic form factors (EFFs) in the space-like and valence quark distribution functions (VQDFs) within the confining Nambu-Jona-Lasinio (NJL) model, which is the chiral effective quark model. These EFFs and VQDFs are the most basic quantities towards our understanding the structure of the $K^{+}$ meson. From the Lagrangian of the model, the dynamical mass of quark or antiquark are generated by interaction with the vacuum state, where the chiral spontaneously symmetry breaking is realized. However the NJL model has a divergent in the momentum loop integral, a specific regularization will therefore be chosen to cure the divergence. In this study the proper time regularization (PTR) scheme is applied. In this regularization the infrared cutoff removes the imaginary part of the loop integral to eliminate the unphysical domain for hadron decay into quarks and simulates confinement. Thus these dynamical features of the kaon in the chiral effective quark model are used as an input into EFFs and PDFs to describe the kaon structure. 

The outline of this paper is as follows. Section~II discuss briefly the kaon structure in the confining NJL model. Section~III shows the results of the elastic form factor of the kaon in the space-like region and the valence quark distribution of the kaon as well as a comparison with the limited experimental data for the kaon, and finally the conclusion is presented in Section~IV.

\begin{center}
\textbf{II. KAON STRUCTURE IN THE CONFINING NJL MODEL}
\end{center}

The three-flavour NJL Lagrangian -- containing only four-fermion interactions -- takes the form~[1]
\begin{align}
\mathcal{L}_{NJL} &= \bar{\psi}(i\slashed{\partial} - \hat{m})\psi + G_\pi \left[ \left( \bar{\psi} \lambda_a \psi \right)^2 - \left( \bar{\psi} \lambda_a \gamma_5 \psi \right)^2 \right] - G_\rho  \left[ \left( \bar{\psi} \lambda_a \gamma^\mu \psi \right)^2 + \left( \bar{\psi} \lambda_a \gamma^\mu \gamma_5 \psi \right)^2 \right],
\label{NJL lagrangian}
\end{align}
where the quark field has the flavour components $\psi = (u, d, s)$, $\hat{m}={\rm diag} (m_u, m_d, m_s)$ denotes the
current quark mass matrix, and $G_{\pi}, G_\rho$ are the four-fermion coupling constants. Based upon Bethe Salpeter equation for antiquark-quark correlations in the confining NJL model using the random phase approximation, the dressed quark masses in the proper time regularization scheme are given by the solution of the gap equation:
\begin{align}
M_q &= m_q + \frac{3  M_q G_\pi}{\pi^2} \int_{\frac{1}{\Lambda_{UV}^{2}}}^{\frac{1}{\Lambda_{IR}^{2}}} d\tau \frac{e^{-\tau M_q^2}}{\tau^2}
\label{BSE equation}
\end{align}
Kaon EFFs in the space-like region and VQDFs have been calculated in the confining NJL model Ref.~[1, 2], where the kaon is realized in the NJL model as quark-antiquark bound state. After finding the BSE solution, we introduce the reduced $t-$matrice in the kaon channel and the bubble diagrams of the kaon. Then the kaon mass is defined by the pole in the corresponding $t-$matrix. The residue at the pole in the $\bar{q}q$ $t-$matrices defines the effective meson quark-quark coupling constant. These quantities are used as input to calculate the EFFs and VQDFs. Using the matrix element of the electromagnetic current for the kaon, the complete results for the kaon form factors -- with dressed $s-$quark-photon vertex  -- reads
\begin{align}
\label{eq:fullpi}
F_{K^{+}}(Q^2) &= F_{1U}(Q^2)\,f^{\ell s}_K(Q^2) - F_{1S}(Q^2)\,f^{s\ell}_K(Q^2),
\end{align}
where the quark sector form factors are defined as in Ref.~[1].

The valence quark distribution functions of the kaon are extracted using the moments. The functions are obtained after applying the Ward identity $S(k) \gamma^{+} S(k) = - \partial S(k)/\partial_{+}$ and introducing the Feynman parametrization. The valence quark distribution functions can then be straightforwardly determined. For the valence quark and anti-quark distributions of the $K^+$ we find:
\begin{align}
\label{eq:valence5}
q_{K} (x) &= \frac{3Z_K}{4\pi^2} \int d\tau e^{-\tau \left[ x(x-1)m_K^2 + x M_s^2 + (1-x) M_\ell^2 \right]} \left[ \frac{1}{\tau} + x(1-x) \left[m_K^2 - (M_\ell -M_s)^2 \right] \right], \nonumber \\
\bar{q}_{K} (x) &= \frac{3Z_K}{4\pi^2} \int d\tau e^{-\tau \left[ x(x-1)m_K^2 + x M_\ell^2 + (1-x) M_s^2 \right]} \left[ \frac{1}{\tau} + x(1-x) \left[m_K^2 - (M_\ell -M_s)^2 \right] \right]. \nonumber \\
\end{align}
Results for the $\pi^+$ are obtained by $M_s \to M_\ell$ and $Z_K \to Z_\pi$, giving the result $u_{\pi^+}(x) = \bar{d}_{\pi^+}(x)$. Moreover, the valence quark distribution functions for the other pseudoscalar mesons can be obtained using flavour symmetries. The 
valence quark distribution functions must satisfy the baryon number:
\begin{align}
\int_0^1 dx \left[ u_K (x) - \bar{u}_K (x) \right] = \int_0^1 dx \left[ \bar{s}_K (x) - s_K (x) \right] = 1,
\end{align}
and momentum sum rules,  
\begin{align}
\int_0^1 dx x \left[ u_K (x) + \bar{u}_K (x) + s_K (x) + \bar{s}_K (x) \right] = 1.
\end{align}
More details of the rules of the baryon number and momentum can be found in our previous work~[1].

\begin{center}
\textbf{III. RESULTS}
\end{center}

Numerical results for the kaon elastic form factors in the space-like region and the quark sector components -- including effects from the dressed quark-photon vertex-- are illustrated in Figs.~1 and~2. We find excellent agreement with the available experimental data from Ref.~[8] and the empirical monopole, $F_K (Q^2) = \left[ 1 +Q^2/\Lambda_{K}^2 \right]^{-1}$ determined by reproducing the charge radius of Ref.~[8]. By multiplying the kaon form factors with charges, as in Fig.~2, we find that the strange quark component becomes dominant among other components in the $K^{+}$ elastic form factor for $Q^2 \geq 1.6\,$GeV$^2$. This is more completely dominant at very large $Q^2$.

Numerical results for the kaon valence quark distribution functions at $Q^2 = 16 \,$GeV$^2$ are shown in Fig.~3 and compared to empirical data for the pion valence quark distribution function from Ref~[9]. We find reasonable agreement over the entire $x$ domain where the data is existed. Our results have been evolved from a model scale of $Q_0^2 = 0.16\,$GeV$^2$ using the next-to-leading order (NLO) DGLAP evolution equations~[10], which was independently calculated in Ref.~[5] in the study of nucleon parton distribution functions. At the model scale we find that the momentum fraction carried by the $u$ and $s$ quarks in the $K^+$ are $\langle x\,u\rangle = 0.42$ and $\langle x\,s\rangle = 0.58$ (at this scale gluons do not carry momentum so these results saturate the momentum sum rule).

The ratio $u_{K^+}(x)/u_{\pi^+}(x)$ is shown in Fig.~4 at $Q^2 = 16\,$GeV$^2$, however this ratio has only a slight $Q^2$ dependence and
in the limit $x \to 1$ is a fixed point in $Q^2$. We find $u_{K^+}/u_{\pi^+} \to 0.434 \simeq M_u^2/M_s^2$ as $x \to 1$, in good agreement with existing data from Ref.~[11]. However, the $x$ dependence differs from much of the data in the valence region. The reason for this discrepancy is not clear, however it may lie with the absence of momentum dependence in standard NJL Bethe-Salpeter vertices~[6, 13], or with the data itself. We note however the correspondence that $u_{K^+}/u_{\pi^+} < 1$ as $x \to 1$ and that $F^u_K(Q^2)/F_\pi^u(Q^2) < 1$, as shown in Fig.~2, for $Q^2 \gg \Lambda_{\text{QCD}}^2$. Fig.~4 presents the ratio $u_{K^+}(x)/s_{K^+}(x)$, which approaches $0.37$ as $x \to 1$. It reveals that flavour breaking effects have a sizable $x$ dependence, being maximum at large $x$ and becoming negligible at small $x$ where effects of perturbation from DGLAP evolution dominate.

\begin{figure}[tbp]
  \centering
  \begin{tikzpicture}
    \node[anchor=south west,inner sep=0] (image) at (0,0)    {\includegraphics[width=0.65\columnwidth]{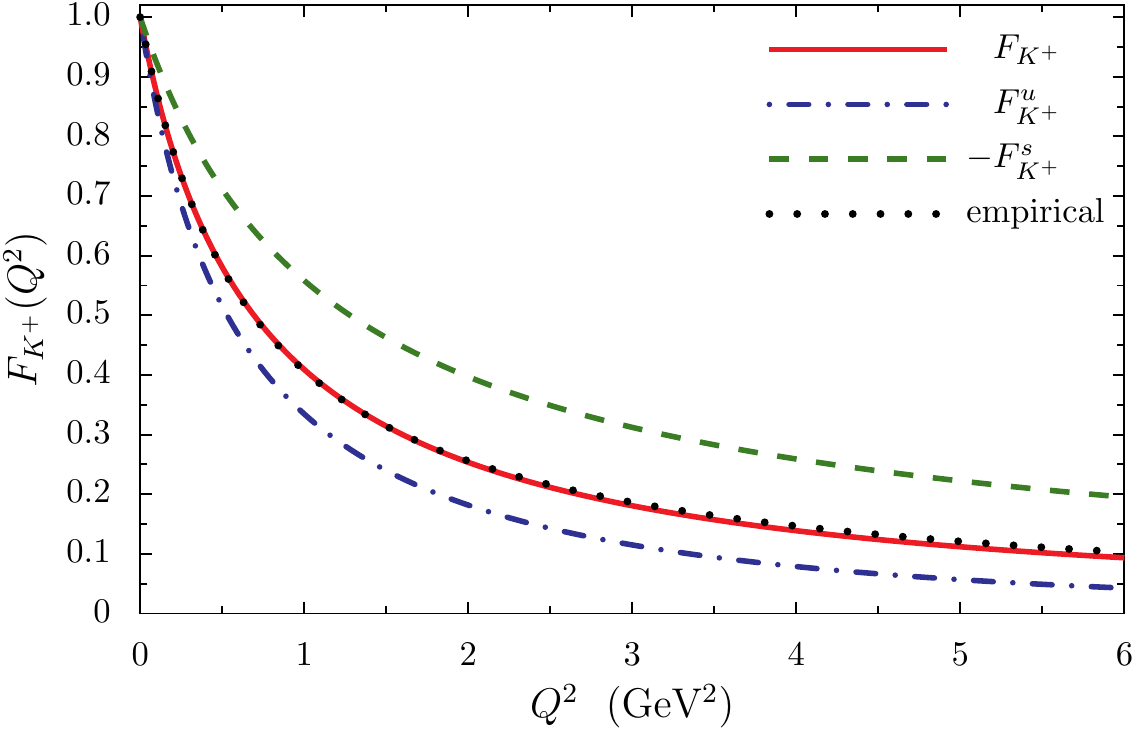}};
    \node[anchor=south west,inner sep=0] (image) at (2.3,4.5){\includegraphics[width=0.28\columnwidth]{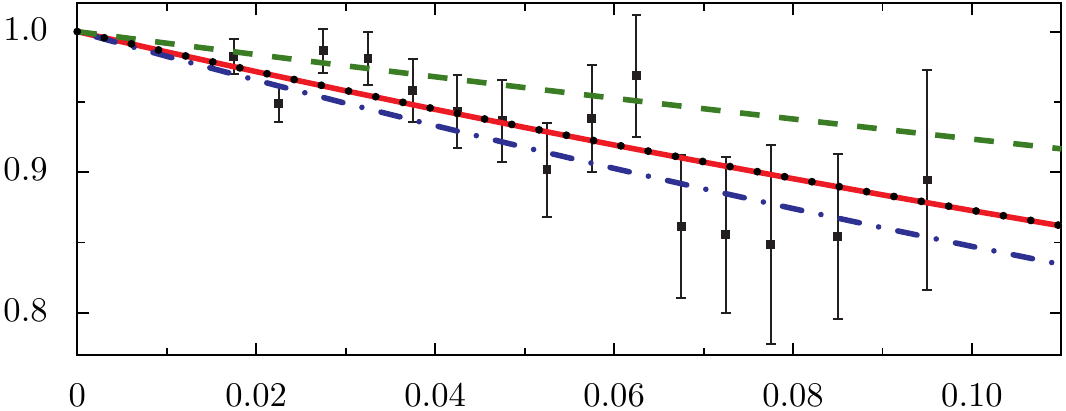}};
  \end{tikzpicture}
\caption{ \small  (Colour online) The $K^+$ form factor (solid line) together with the up 
(dashed-dotted line) and strange (dashed line) quark sector contributions. The dotted-line 
is the fit to data using the form $F_K(Q^2) = [1 + Q^2/\Lambda_K^2]^{-1}$, giving $\Lambda^2_K = 0.687\,$GeV$^2$, and
the insert compares our results with existing data taken from Ref.~[8]. }
\label{fig:kaonff}
\end{figure}

\begin{figure}[tbp]
\centering
\includegraphics[width=0.65\columnwidth]{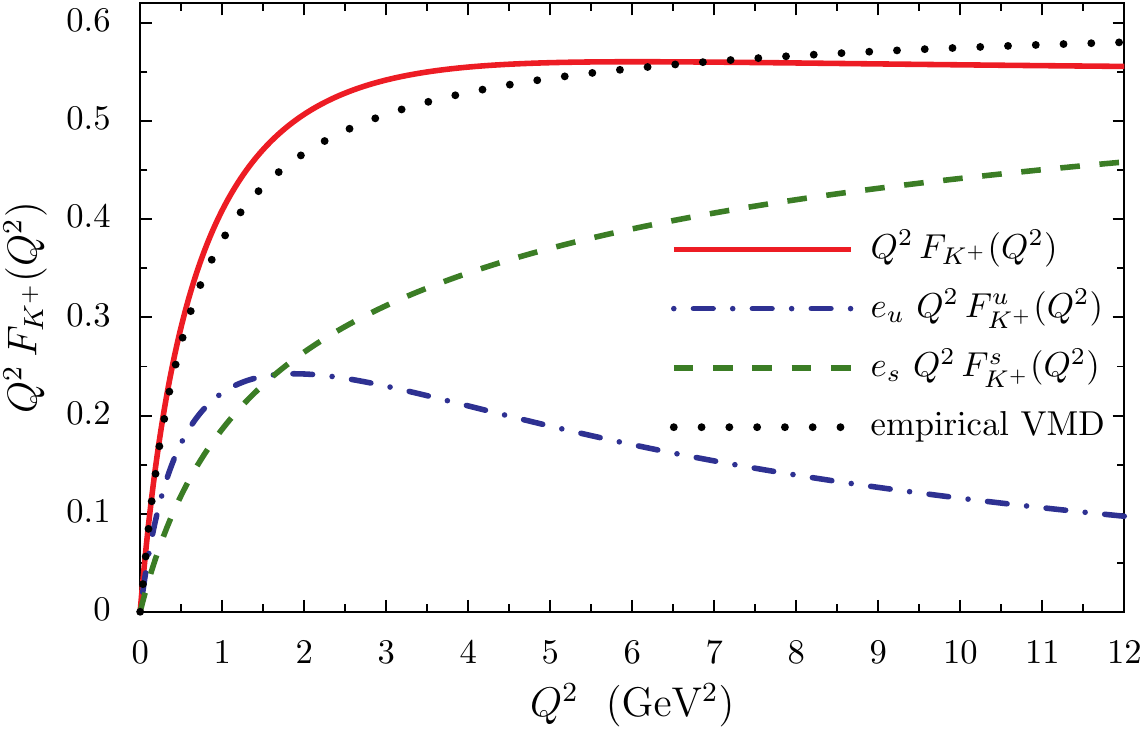}
\caption{  \small  (Colour online) Results for $Q^2\,F_{K^{+}} (Q^2)$ together with the charge-weighted 
quark-sector contributions and the empirical result obtained from Ref.~[8].
This result clearly illustrates that the $s-$quark dominates the form factor at large $Q^2$.}
\label{fig:Q2kaonff} 
\end{figure}

\begin{figure}[tbp]
\centering
\includegraphics[width=0.65\columnwidth]{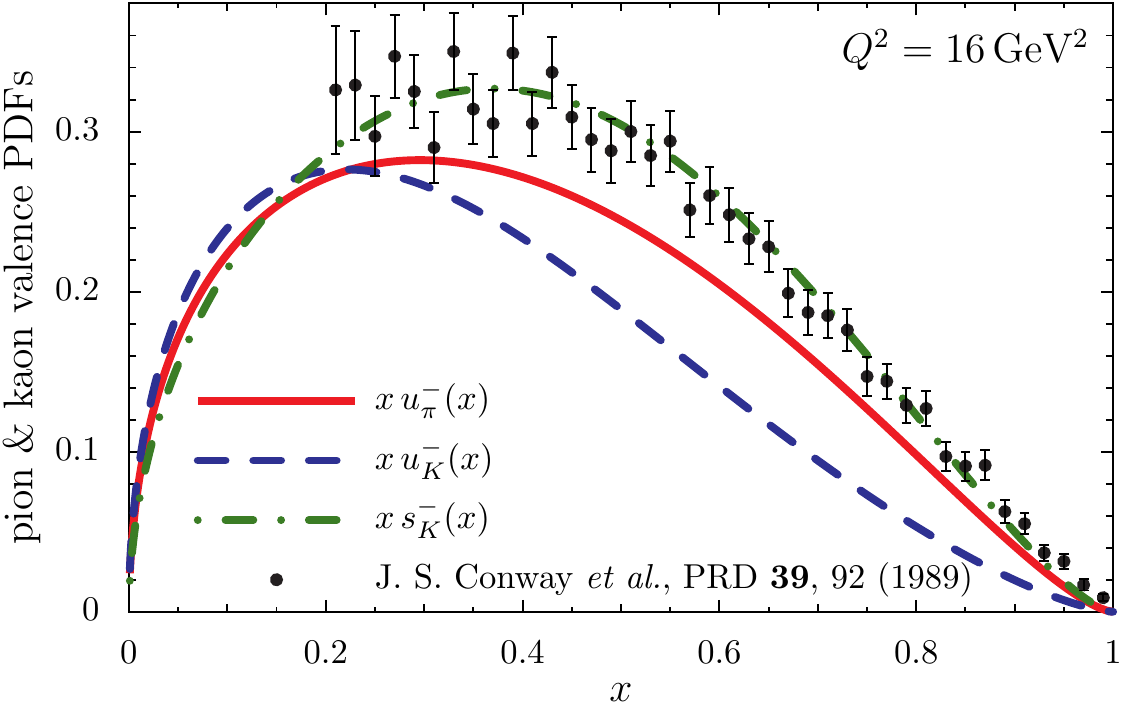}
\caption{(Colour online) Results for the valence quark distributions 
of the $\pi^{+}$ and $K^{+}$, evolved from the model scale using the NLO DGLAP
equations~[10]. The solid line represents the valence $u$ or $\bar{d}$ PDF
in the $\pi^{+}$, the dot-dashed line is the valence $\bar{s}$ quark and the 
dashed line the valence $u$ quark in the $K^{+}$. The experimental data 
are taken from Ref.~[9].}
\label{fig:pdfs} 
\end{figure}

\begin{figure}[tbp] 
\centering\includegraphics[width=0.65\columnwidth]{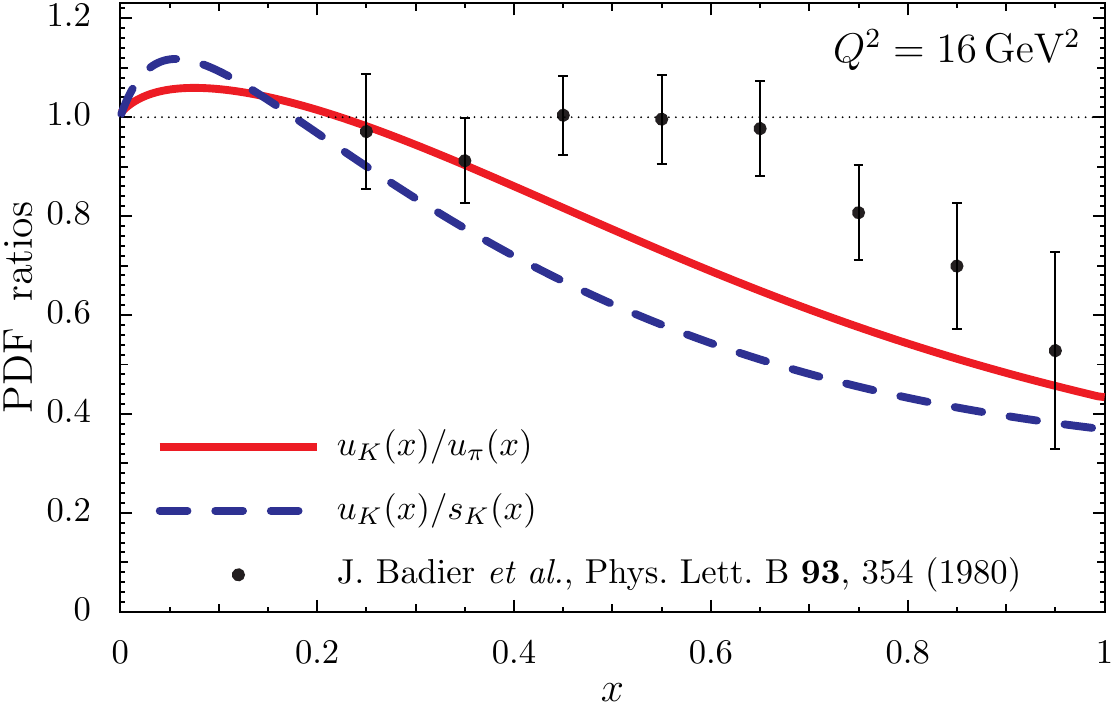}
\caption{(Colour online) The solid line gives the ratio of the $u-$quark distribution of the kaon to the $u-$quark distribution of the pion, after NLO evolution to $Q^2$ = 16 $\rm{GeV}^2$. The dashed line gives the ratio of the $u$ quark to $s$ quark distributions in the kaon at $Q^2$ = 16 $\rm{GeV}^2$. The experimental data 
are taken from Ref.~[11].}
\label{fig:pdfratio} 
\end{figure}

\begin{center}
\textbf{IV. CONCLUSION}
\end{center}

We have used the confining NJL model to calculate the electromagnetic form factors and VQDFs of the kaon. 
We have included the effect of dressed vertex through vector meson like correlations in the $t$-channel, which do not contribute to the VQDFs.

This work has produced several remarkable results. In comparison with the experimental data and empirical formula, the total kaon form factor agrees very well with the limited existing data. The effects of the strange quark mass on the VQDFs is less spectacular. In Fig.~3 we saw that the strange quark VQDFs in the $K^+$ is considerably enhanced over that of the $u$-quark in the valence region. Most importantly, as we see in Fig.~4, the empirical suppression of $u_{K^+}$ compared 
with $u_{\pi^+}$ is rather well described. 

Experimental data on the kaon form factors and valence quark distribution functions  are sparse. In the near future the new data for kaon are expected from CLAS, JPARC and COMPASS experiment at CERN as well as the future EIC (Electron Ion Collider) experiment. The result of this work will be tested using those new experimental data to understand some issues on structure of the kaon such as a flavor symmetry breaking~[1, 4, 5] as well as the gluon distribution~[12, 13] in the kaon. On the other hand, the comparison between our model prediction and experimental data will lead us to new understanding of the kaon structure and QCD as underlying theory.

\begin{center}
\textbf{ACKNOWLEDGMENTS}
\end{center}
This work was supported by the Young Scientist Training program of the Asia Pacific Center for Theoretical Physics, Pohang, South Korea. 
\\

\begin{center}
\textbf{REFERENCES}
\end{center}

\begin{spacing}{0.5}
\begin{enumerate}
\item P.~T.~P.~Hutauruk, I.~C.~Cloet and A.~W.~Thomas, Flavor dependence of the pion and kaon form factors and parton distribution functions,\textit{Phys.\ Rev.\ C {\bf 94}}, 035201 (2016).
\item Y.~Ninomiya, W.~Bentz and I.~C.~Cloet, Dressed Quark Mass Dependence of Pion and Kaon Form Factors, \textit{Phys.\ Rev.\ C {\bf 91}}, 025202 (2015).
\item A.~F.~Krutov, S.~V.~Troitsky and V.~E.~Troitsky, The $K$-meson form factor and charge radius: linking low-energy data to future Jefferson Laboratory measurements,\textit{ Eur.\ Phys.\ J.\ C {\bf 77}}, 464 (2017).
\item T.~Horn, Meson Form Factors and Deep Exclusive Meson Production Experiments, \textit{EPJ Web Conf.\  {\bf 137}}, 05005 (2017).
\item I.~C.~Cloet, W.~Bentz and A.~W.~Thomas, Role of diquark correlations and the pion cloud in nucleon elastic form factors, \textit{Phys.\ Rev.\ C {\bf 90}}, 045202 (2014).
\item T.~Nguyen, A.~Bashir, C.~D.~Roberts and P.~C.~Tandy, Pion and kaon valence-quark parton distribution functions, \textit{Phys.\ Rev.\ C {\bf 83}}, 062201 (2011).
\item R.~K.~Ellis, Quantum Chromodynamics and Deep Inelastic Scattering, \textit{Adv.\ Ser.\ Direct.\ High Energy Phys.\  {\bf 26}}, 61 (2016). 
\item S.~R.~Amendolia {\it et al.} [NA7 Collaboration], A Measurement of the Space - Like Pion Electromagnetic Form-Factor, \textit{Nucl.\ Phys.\ B {\bf 277}}, 168 (1986).
\item J.~S.~Conway {\it et al.}, Experimental Study of Muon Pairs Produced by 252-GeV Pions on Tungsten, \textit{Phys.\ Rev.\ D {\bf 39}}, 92 (1989).
\item M.~Miyama and S.~Kumano, Numerical solution of Q**2 evolution equations in a brute force method, \textit{Comput.\ Phys.\ Commun.\  {\bf 94}}, 185 (1996).
\item J.~Badier {\it et al.} [Saclay-CERN-College de France-Ecole Poly-Orsay Collaboration], Measurement of the $K^- / \pi^-$ Structure Function Ratio Using the {Drell-Yan} Process, \textit{Phys.\ Lett.\ B {\bf 93}}, 354 (1980).
\item J.~C.~Peng, W.~C.~Chang, S.~Platchkov and T.~Sawada, Valence Quark and Gluon Distributions of Kaon from J/Psi Production, \textit{ arXiv:1711.00839 [hep-ph]}.
\item C.~Chen, L.~Chang, C.~D.~Roberts, S.~Wan and H.~S.~Zong, Valence-quark distribution functions in the kaon and pion, \textit{Phys.\ Rev.\ D {\bf 93}}, 074021 (2016).
\end{enumerate}
\end{spacing}
\end{document}